\documentclass[12pt]{article}

\usepackage{amsmath}
\usepackage{amssymb}

\newcommand{\tmop}[1]{\operatorname{#1}}
\newcommand{\beq}{\begin{equation}}
\newcommand{\eeq}{\end{equation}}
\begin{document}
\begin{titlepage}
\begin{flushright}
YITP-SB-04-45  \\
{\tt hep-th/0408188}\\
\end{flushright}
\vskip 22mm
\begin{center}
{\huge {\bf On Calabi-Yau supermanifolds}}
\vskip 10mm
{\bf Martin Ro\v{c}ek}\\
{\em C.N. Yang Institute for Theoretical Physics}\\
{\em SUNY, Stony Brook, NY 11794-3840, USA}\\ 
{\tt rocek@insti.physics.sunysb.edu}
\vskip 6mm
{\bf Neal Wadhwa}\\
{\em Ward Melville High School}
\end{center}
\vskip .2in

\begin{center} {\bf ABSTRACT } \end{center}
\begin{quotation}\noindent
We prove that a K\"ahler supermetric on a supermanifold with one complex 
fermionic dimension admits a super Ricci-flat supermetric if and only if the bosonic metric 
has vanishing scalar curvature. As a 
corollary, it follows that Yau's theorem does not hold for supermanifolds.
\end{quotation}
\vfill
%%%%%%%%%%%%%%%%%%
\end{titlepage}
\newpage

Calabi\cite{calabi} proposed that if a K\"ahler manifold has vanishing first
Chern class, that is, the Ricci-form obeys $R_{i \bar{j}} ( g ) =
\partial_i  \bar{v}_j - \bar{\partial}_{j} v_i$ for a globally defined
1-form $v$, or, equivalently, a complex $n$-dimensional K\"ahler manifold
has a globally defined holomorphic top form $\Omega_{i_1\dots i_n}$, 
then there exists a unique metric $g'$ which is a smooth
deformation of $g$ and obeys $R_{i \bar{j}} ( g' ) = 0$. Yau\cite{yau} proved this
theorem for ordinary manifolds. 

Recently, there has been a lot of interest in Calabi-Yau supermanifolds \cite{cysm};
though these papers use only the topological properties of such spaces,
it is interesting to ask whether they also admit Ricci-flat supermetrics.
This paper studies the generalization
of Calabi's conjecture to supermanifolds with one complex fermionic dimension. 
We find that such a K\"ahler supermanifold admits a
Ricci-flat supermetric if and only if the bosonic metric has vanishing scalar curvature. For a
given scalar-flat bosonic K\"ahler metric with K\"ahler potential $K_{Bose}$, 
the super-extension is unique, and
has the super K\"ahler potential:
\beq
K(z^i,\bar z^j,\theta,\bar \theta)=K_{Bose}(z^i,\bar z^j)+
\det\!\!\left(\!\! \frac{\partial^2 }{\partial z^i\partial\bar z^j}K_{Bose}\!\!\right)\!\theta\bar\theta~.
\eeq
As complex projective spaces do not admit scalar-flat metrics, but do admit super Calabi-Yau
extensions with one fermionic dimension, 
it follows that Yau's theorem does not hold for supermanifolds.

A supermanifold is a generalization of a usual manifold 
with {\em fermionic} as well as bosonic 
coordinates\footnote{More rigorous and technical definitions 
can be found in the literature, see, {\it e.g.,} \cite{dewitt}, 
but this simple treatment suffices for our results.}.
The bosonic coordinates are ordinary numbers, 
whereas the fermionic coordinates are grassmann numbers.
Grassmann numbers are odd elements of a grassmann algebra and anticommute:
$\theta^1 \theta^2 = - \theta^2 \theta^1$ and $\theta^1 \theta^1 = 0$. 

On bosonic K\"ahler manifolds, the Ricci tensor 
\beq
R_{i \bar{j}} = (
\ln \det ( g ) )_{, i \bar{j}}~.
\eeq
For this to vanish, $\ln \det ( g ) )$ (locally) must be the real part of a holomorphic
function, and hence $\det ( g ) )=|f(z)|^2$
for some holomorphic $f(z)$. This can always be absorbed by a holomorphic coordinate
transformation, and hence a K\"ahler manifold is Ricci-flat if its K\"ahler potential $K$ obeys the 
Monge-Amp\`ere equation
\beq
\det ( g )\equiv \det(K_{,i\bar j})=1~.
\eeq
On supermanifolds, because elements of $g$ contain
grassmann numbers, the determinant is not well defined and a new definition of the
determinant is needed. For any nondegenerate supermatrix
\beq
g = \left(\begin{array}{cc}
  A & B\\
  C & D
\end{array}\right)~,
\eeq
where $A$ and $D$ are bosonic and $B$ and $C$ are
fermionic, 
\beq
\tmop{sdet} ( g ) \equiv \frac{\det ( A )}{\det ( D - \tmop{CA}^{- 1} B
)} = \frac{\det ( A - BD^{- 1} C )}{\det ( D )}~.
\eeq
For arbitrary supermatrices $X,Y$, this definition is consistent with the 
basic relation $\tmop{sdet}(XY)=\tmop{sdet}(X)\tmop{sdet}(Y)$.
In addition, the supertrace is defined as 
\begin{equation}\tmop{str} (g) = \tmop {tr} (A)-\tmop {tr} (D)~,
\end{equation} 
which is consistent with $\tmop{str}(XY)=\tmop{str}(YX)$.
These two definitions imply an identity that is useful in simplifying 
expressions that use grassmann numbers: 
\beq
\label{lnsdet}
\tmop{ln\,sdet}(g)=\tmop{str\,ln}(g)
\eeq
Simple examples\cite{sethi} of K\"ahler supermanifolds are provided 
by superprojective spaces, $SP(m|n)$.
These can be described in terms of $m+n+1$ homogeneous coordinates: 
\beq
(z^1,z^2,...,z^{m+1}|\theta^1,...,\theta^n)
\eeq
related by the equivalence relations $z^{i}  \sim \lambda z^i$ and 
$\theta^{i} \sim \lambda \theta^i$. There are $m+1$ coordinate patches 
where $z^i\neq0$ in the $i$-th coordinate patch. In the $i$-th patch, we can 
introduce inhomogeneous coordinates $\tilde z^j=\frac{z^j}{z^i}$. 
Other examples include weighted superprojective space, 
$WSP(k_1\ldots k_{m+1}|l_1,\ldots,l_n)$; the coordinates are identified 
under the equivalence relations $z^i\sim \lambda^{k_i} z^i$ 
and $\theta^i\sim\lambda^{ l_i}\theta^i$.
A direct calculation of the Ricci-form of the standard Fubini-Study metric reveals that 
$SP(m|m+1)$ are Calabi-Yau and have a vanishing Ricci-form, whereas 
$WSP(1\ldots1|m)$ are Calabi-Yau but have
a nonvanishing Ricci-form (see below).

We now show that for an arbitrary K\"ahler space with only one complex fermionic coordinate, 
$R_{i \bar{j}} = 0$ implies that the bosonic part of the K\"ahler potential
yields a space with a Ricci scalar $s=0$. Consider an arbitrary super K\"ahler potential $K$ 
on a supermanifold $\mathcal{M}(m|1)$ with one complex fermionic coordinate $\theta$ 
and $m$ bosonic coordinates. The super K\"ahler potential can be written as $K = f^0 + f^1 \theta
\bar{\theta}$.  We use the convention that holomorphic derivatives are taken from the 
left and anti-holomorphic derviatives are take from the right. 
The supermetric $\mathbf{g}$ is the block matrix 
\beq
\mathbf{g}= 
\left(\begin{array}{cc}
 f^{0_{}}_{, i \bar{j}} + f^1_{,i \bar{j}} \theta \bar{\theta} & f^1_{, i}
 \theta\\ & \\
 f^1_{, \bar{j}}  \bar{\theta} & f^1
\end{array}\right)~.
\eeq 
Its superdeterminant is 
\begin{eqnarray}
\tmop{sdet}(\mathbf{g}) &=&
\frac{\det [ f^0_{, i\bar{j}} + ( f^1_{,i \bar{j}} - f^1_{, i} f^1_{, \bar{j}} / f^1 ) \theta
\bar{\theta} ]}{f^1}\nonumber\\ && \nonumber\\
&=&\frac{\det ( f^0_{, i\bar{j}})}{f^1}\det\!\!\left[\!\delta^k_i+g^{k\bar j}\!\! 
\left(\!\! f^1_{,i \bar{j}} - \frac{f^1_{, i} f^1_{, \bar{j}}}{f^1}\!\right)\!\!\theta
\bar{\theta}\! \right]~,
\end{eqnarray}
where $ g^{ i \bar{j}}\equiv(f^0_{, i
\bar{j}})^{-1}$ is the inverse metric of the bosonic manifold.
Using the identity (\ref{lnsdet}), we can rewrite this as:
\beq
\tmop{sdet}(\mathbf{g}) =\frac{\det ( f^0_{, i\bar{j}})}{f^1}\!\!\left[1+g^{i\bar j}\!\! 
\left(\!\! f^1_{,i \bar{j}} - \frac{f^1_{, i} f^1_{, \bar{j}}}{f^1}\!\right)\!\!\theta
\bar{\theta}\! \right]~.
\eeq
On a super Ricci-flat manifold, the superdeterminant can be chosen to be $1$.
The $\theta$-independent term of $\tmop{sdet}(\mathbf{g})=1$ implies
\beq 
f^1 = \det ( f^0_{, i \bar{j}} )~.
\label{f1}
\eeq
The remaining term must vanish on a super Ricci-flat K\"ahler manifold. This implies
\beq
g^{i\bar j}\!\! \left(\!\! f^1_{,i \bar{j}} - \frac{f^1_{, i} f^1_{, \bar{j}}}{f^1}\!\right)=
f^1g^{i\bar j}[\ln(f^1)]_{,i\bar j}=0~.
\eeq
Substituting (\ref{f1}) implies
\beq
g^{{ i \bar{j}}}
\tmop{ln} \,\tmop{det}(f^0_{,l\bar{k}})_{, i \bar{j}}\equiv g^{ i \bar{j}}R_{i\bar{j}}=0~,
\eeq
which is precisely the Ricci scalar of the bosonic space with K\"ahler potential $f^0$. This proves our
main result: a K\"ahler supermanifold with one complex fermionic dimension admits a super Ricci-flat
extension if and only if the bosonic K\"ahler manifold that it is based on has vanishing scalar curvature 
$s$.  Many such bosonic manifolds are known and have been studied; see, {\it e.g.,} \cite{lebrun}; such
spaces all admit supermanifolds with Calabi-Yau supermetrics. A simple example is the space
$\mathbb{CP}^1\times \Sigma$, where $\Sigma$ is a Riemann surface with a metric with constant
curvature chosen so that the total scalar curvature vanishes. The super Ricci-flat K\"ahler potential on 
such a space is
\beq
K=\tmop{ln}(1+z_1\bar{z}_1)-\tmop{ln}(1-z_2\bar{z}_2)+
\frac{\theta\bar\theta}{(1+z_1\bar{z}_1)^2(1-z_2\bar{z}_2)^2}~.
\eeq
There are many other $s=0$ metrics which can be studied this way.

A corollary of our result is that there are many K\"ahler supermanifolds with vanishing first Chern class
that do not admit super Ricci-flat supermetrics, thus proving that Yau's theorem does not apply to
supermanifolds. Clearly, since no projective space admits an $s=0$ metric, no supermanifold with 
one complex fermionic coordinate that is based on projective space admits a super Ricci-flat supermetric. 
To find our counterexample, it suffices to to prove that such supermanifolds may have 
vanishing first Chern class.

We now consider the explicit example $WSP(1,1|2)$. The superprojective space $WSP(1,1|2)$ 
has a bosonic base which is just $\mathbb{CP}^1$, and $\ln\det(\mathbf{g})$ of the Fubini-Study 
supermetric is the {\em globally defined} scalar,$\frac{\theta \bar{\theta}}{( 1 + z \bar{z} )^2}$. 
The gradient of this scalar is a globally defined vector that fulfills the conditions of the super 
Calabi-Yau conjecture. Equivalently, the top form $dz\wedge d\theta$
is a globally defined holomorphic top-form (the superdeterminant of the coordinate transformation
$z\to-1/z,~\theta\to\theta/z^2$ between the two patches that cover $\mathbb{CP}^1$ is 1).
As the bosonic part of $WSP(1,1|2)$, $\mathbb{CP}^1$, has
no metric with Ricci scalar $s=0$, this space does not satisfy the super Calabi-Yau conjecture. This
result can be generalized to $WSP(\,\underbrace{1,...,1}_m|m)$. These spaces have a globally defined 
vectors on them that fulfill the conditions of the super Calabi-Yau conjecture or equivlently they 
have globally-defined holomorphic top-forms that exist in every coordinate patch. 
In \cite{sethi}, it is observed
that $WSP(1,1|2)$ appears to violate the super Calabi-Yau conjecture, though no explicit proof is
given, and it is conjectured that $WSP(1,...,1|m)$ for $m>2$ will satisfy the conjecture; here we have
shown that no $WSP(1,...,1|m)$ admits a super Ricci-flat supermetric. 

\bigskip
\noindent{Acknowledgement: We are happy to thank C. LeBrun, S. Sethi, C. Vafa, E. Witten, 
and S.T. Yau for encouragement and helpful comments, and the Second Simons Workshop
in Physics and Mathematics for a stimulating environment. The work of MR was supported 
in part by NSF grant no.~PHY-0354776.}

\end{document}